\documentclass[12pt,a4paper]{article}
\pdfoutput=1
\usepackage[latin1]{inputenc}

\usepackage{amsmath}
\usepackage{color}
\usepackage{amsfonts}
\usepackage{amssymb}
\usepackage{graphicx}
\usepackage{geometry}
\usepackage{amssymb,epsfig,subfigure}
\usepackage{hyperref}
\usepackage{comment}
\usepackage[font=footnotesize]{caption}

\makeatletter
\renewcommand\section{\@startsection {section}{1}{\z@}%
                                 {-3.5ex \@plus -1ex \@minus -.2ex}%nn
                                   {2.3ex \@plus.2ex}%
                                   {\normalfont\large\bfseries}}
\renewcommand\subsection{\@startsection{subsection}{2}{\z@}%
                                   {-3.25ex\@plus -1ex \@minus -.2ex}%
                                     {1.5ex \@plus .2ex}%
                                     {\normalfont\bfseries}}
\renewcommand\subsubsection{\@startsection{subsubsection}{3}{\z@}%
                                   {-3.25ex\@plus -1ex \@minus -.2ex}%
                                     {1.5ex \@plus .2ex}%
                                     {\normalfont\itshape}}
\makeatother

\def\pplogo{\vbox{\kern-\headheight\kern -29pt
\halign{##&##\hfil\cr&{\ppnumber}\cr\rule{0pt}{2.5ex}&\ppdate\cr}}}
\makeatletter
\def\ps@firstpage{\ps@empty \def\@oddhead{\hss\pplogo}%
  \let\@evenhead\@oddhead % in case an article starts on a left-hand page
}%      The only change in \maketitle is \thispagestyle{firstpage} instead of 
\def\maketitle{\par
 \begingroup
 \def\thefootnote{\fnsymbol{footnote}}
 \def\@makefnmark{\hbox{$^{\@thefnmark}$\hss}}
 \if@twocolumn
 \twocolumn[\@maketitle]
 \else \newpage
 \global\@topnum\z@ \@maketitle \fi\thispagestyle{firstpage}\@thanks
 \endgroup
 \setcounter{footnote}{0}
 \let\maketitle\relax
 \let\@maketitle\relax
 \gdef\@thanks{}\gdef\@author{}\gdef\@title{}\let\thanks\relax}
\makeatother

\numberwithin{equation}{section}

\newcommand\eea{\end{eqnarray}}
\newcommand\bea{\begin{eqnarray}}

\def\beq{\begin{equation}}
\def\eeq{\end{equation}}

\newcommand{\be}{\begin{equation}}
\newcommand{\ee}{\end{equation}}
\newcommand{\ba}{\begin{align}}
\newcommand{\ea}{\end{align}}
\newcommand{\bg}{\begin{gather}}
\newcommand{\eg}{\end{gather}}
\newcommand{\bseq}{\begin{subequations}}
\newcommand{\eseq}{\end{subequations}}

\renewcommand{\t}{\tilde}

\newcommand{\tr}{{\rm tr}}
\newcommand{\mc}{\mathcal}

\begin{document}
\setcounter{page}0
\def\ppnumber{\vbox{\baselineskip14pt
%\hbox{hep-th/0000000}
}}
\def\ppdate{
%\footnotesize{SU/ITP-14/XX}
} \date{}

\author{Horacio Casini, Eduardo Test\'e, Gonzalo Torroba\\
[7mm] \\
{\normalsize \it Centro At\'omico Bariloche and CONICET}\\
{\normalsize \it S.C. de Bariloche, R\'io Negro, R8402AGP, Argentina}
}

\bigskip
\title{\bf  Relative entropy and the RG flow
\vskip 0.5cm}
\maketitle

\begin{abstract}
We consider the relative entropy between vacuum states of two different theories: a conformal field theory (CFT), and the CFT perturbed by a relevant operator. By restricting both states to the null Cauchy surface in the causal domain of a sphere, we make the relative entropy equal to the difference of entanglement entropies. As a result, this difference has the positivity and monotonicity properties of relative entropy. From this it follows a simple alternative proof of the $c$-theorem in $d=2$ space-time dimensions and, for $d>2$, the proof that the coefficient of the area term in the entanglement entropy decreases along the renormalization group (RG) flow between fixed points. We comment on the regimes of convergence of relative entropy, depending on the space-time dimensions and the conformal dimension $\Delta$ of the perturbation that triggers the RG flow.
\end{abstract}
\bigskip

\newpage

\tableofcontents

\vskip 1cm

%%%%%%%%%%%%%%%%%%%%%%%%%%%%%%%%%%%%%%%%%%%%%%%%%%%%%%%%%%%%%%%%%
%%%%%%%%%%%%%%%%%%%%%%%%%%%%%%%%%%%%%%%%%%%%%%%%%%%%%%%%%%%%%%%%%
%%%%%%%%%%%%%%%%%%%%%%%%%%%%%%%%%%%%%%%%%%%%%%%%%%%%%%%%%%%%%%%%%
%%%%%%%%%%%%%%%%%%%%%%%%%%%%%%%%%%%%%%%%%%%%%%%%%%%%%%%%%%%%%%%%%
\section{Introduction}\label{sec:intro}

The renormalization group (RG) flow describes how physics changes with scale in a quantum field theory (QFT). In recent years, interesting connections of these flows with quantum information theory (QIT) have been discovered. A universal term in the vacuum entanglement entropy (EE) was shown to decrease monotonically along the RG for space-time dimensions $d=2,3$ \cite{Casini:2004bw, Casini:2012ei, Casini:2015woa}. This gives an alternative proof of the c-theorem in $d=2$ \cite{Zamolodchikov:1986gt} and a proof of the F-theorem in $d=3$ \cite{Myers:2010xs,Jafferis:2011zi}. In addition to unitarity and Lorentz covariance of the QFT, the key property of these proofs is strong subadditivity of entanglement entropy. Holographically, the monotonicity of the RG flow is related to the null energy condition in the bulk \cite{Myers:2010xs, Liu:2012eea}. More generally, the fine-grained RG flow in terms of tensor networks \cite{Vidal:2007hda} has been proposed as a description of the spatial structure of the holographic gravity dual \cite{Swingle:2009bg}.   

A natural information theory tool to study changes between states is the relative entropy. This meassures distinguishability between different states in a precise operational way \cite{Vedral}. In the context of the renormalization group flows a natural idea is to use relative entropy to quantify how a theory (or its vacuum state) gets modified as we change the scale.\footnote{Previous steps in this direction include \cite{Balasubramanian:2014bfa, Gaite:1996vk}, who studied the classical relative entropy between the probability distributions defined by the Euclidean path integrals as a measure of distinguishability between theories. A change in the Lagrangian that induces the RG flow produces a change in the state associated to the path integral probability distribution.} 
In this work we consider quantum relative entropies in real time, between vacuum states of two theories reduced to certain regions, and look at the consequences of positivity and monotonicity of relative entropy. We follow the steps of the recent work \cite{Casini:2016fgb}, where relative entropy was shown to lead to a simple proof of the g-theorem for $d=2$ conformal field theories (CFT) in a space with a boundary at $x=0$.    

Evidently, not every pair of vacuum states of two different theories can be compared through the relative entropy. Different theories, i.e. containing  one  and two free scalar fields respectively, usually live in different Hilbert spaces, and there is no natural meaning in taking a relative entropy in this case. In order to compute a relative entropy, we need that (at least in presence of a physical UV cutoff such as a lattice) the microscopic constituents of the two models be the same. For this reason, we will study theories with the same UV fixed point, where this can in principle be achieved. More precisely, we will fix as a reference state the UV conformal fixed point itself, and study the relative entropy with another state arising from the CFT by perturbing it with a relevant operator. We will argue that relative entropy gives a useful notion of statistical distance between these theories, and is well-suited for capturing global properties of RG flows.

Relative entropy is notoriously efficient in distinguishing states. It essentially takes into account all fine grained information about the states. In our setup this is reflected in the possible presence of divergences. In order to get definite results for RG flows, we need to avoid these divergences and prevent the relative entropy from distinguishing the states too much.

Divergences may be of UV origin, due to the fact that even if the two theories we consider approach each other at short distances, the correlators of the deformed theory do not converge to the ones of the CFT fast enough to make the relative entropy finite. We will find a range of the conformal dimension $\Delta$ of the perturbation that triggers the RG flow where relative entropy is free from UV divergences. 

There are also divergences of infrared origin, coming from the difference between the states that pile up for large distances. In fact, if we take the two full vacuum states relative entropy will always be divergent as they correspond to two different pure states. However, this problem is circumvented by looking at the states reduced to a finite region in space. The size $R$ of the region will be the parameter with which we can look at the RG scale. In general, we find that relative entropy increases super-volumetrically as $R^{d}$ due to the contribution of the modular Hamiltonian. Following \cite{Casini:2016fgb}, we will then compare the states on a null surface. This effectively reduces the relative entropy to terms increasing like the area $\sim R^{d-2}$, giving direct information on the entanglement entropy and aspects of its RG flow. 

The main result is a new proof of the $c$-theorem in $d=2$, that extends to higher dimensions $d>2$ as a statement about the renormalization of the area term in entanglement entropy. This is shown to be always decreasing between fixed points, but there is a restricted window of conformal dimensions  $\Delta<(d+2)/2$ where the change is finite. This is parallel to studies of the renormalization of the Newton constant \cite{Adler:1982ri, Zee:1980sj, Casini:2014yca, Casini:2015ffa}. 

The expression in terms of relative entropy gives a more transparent information-theoretic interpretation to these RG monotonicity results. The $c$-theorem is equivalent to the following QIT statement: The vacuum $\rho_1$ of an RG-running theory can be distinguished (using the relative entropy measure) from the vacuum $\rho_0$ of the UV fixed point, compared on the null Cauchy surface of a sphere of radius $R$, by the amount  
\begin{equation}\label{QIT} 
S(\rho^1|\rho^0) \approx \frac{c_{UV}-c_{IR}}{3} \log(m R)\ ,
\end{equation}
for radius $R$ bigger than the scale $m$ characterizing the RG flow; $c_{UV}$ and $c_{IR}$ are the central charges of the UV and IR fixed points. Then the central charge difference $c_{UV} - c_{IR}$ controls the distinguishability, or statistical distance, between the two theories. The $c$-theorem then amounts to positivity and monotonicity of the relative entropy, and can be explained as due to the increased distinguishability of two states as we increase the algebra of operators that are available to probe them. In higher dimensions, we prove a similar inequality for the difference in the EE area terms of the two theories.

The work is organized as follows. In Sec. \ref{sec:relative} we study relative entropy for the vacuum states of two theories, its dependence on the Cauchy surface where the states are compared, and whether this relative entropy is finite or UV divergent. In Sec. \ref{sec:consequences} we study the consequences of positivity and monotonicity of relative entropy evaluated on a null Cauchy surface. We prove the $c$-theorem in $d=2$ and the area theorem for the entanglement entropy in $d>2$. In Sec. \ref{sec:concl} we discuss the results. Finally, the Appendix describes explicit computations for free fields.

%%%%%%%%%%%%%%%%%%%%%%%%%%%%%%%%%%%%%%%%%%%%%%%%%%%%%%%%%%
%%%%%%%%%%%%%%%%%%%%%%%%%%%%%%%%%%%%%%%%%%%%%%%%%%%%%%%%%%
%%%%%%%%%%%%%%%%%%%%%%%%%%%%%%%%%%%%%%%%%%%%%%%%%%%%%%%%%%
%%%%%%%%%%%%%%%%%%%%%%%%%%%%%%%%%%%%%%%%%%%%%%%%%%%%%%%%%%
\section{Relative entropy for states of different theories}\label{sec:relative}

The relative entropy between two density matrices $\rho^0$ and $\rho^1$ is defined by
\begin{equation}\label{rel} 
S(\rho^1|\rho^0) = \text{Tr} \left( \rho^1 \log \rho^1 - \rho^1 \log \rho^0   \right)\ .
\end{equation}
We are interested in the relative entropy of the vacuum states of two theories, reduced to certain surfaces. The surfaces are usual spatial, but we will also consider the null case. The two theories are denoted by ${\cal T}_0$, and ${\cal T}_1$. We are going to take ${\cal T}_0$ as a CFT and ${\cal T}_1$ is obtained by perturbing ${\cal T}_0$ with a relevant deformation, starting an RG flow:
\be\label{eq:RG1}
S_1= S_0+ \int d^d x\,g\,\mathcal O(x)\,.
\ee
The scaling dimension of the operator $\mathcal O$ at the fixed point $g=0$ is denoted by $\Delta$; the perturbation is relevant for $\Delta<d$.
This construction ensures that ${\cal T}_0$ and ${\cal T}_1$ have the same operator content in the UV.    
As these states belong to two different theories, they are evolved in time with two different Hamiltonians. Hence, we have to be more specific on the instant of time when we compare the states, because they will undergo different unitary evolutions, and as a consequence relative entropy will depend on time.

As shown in \cite{Casini:2016fgb} for the simpler setup of the $g$-theorem, the dependence of relative entropy on the Cauchy surface can be exploited to reduce (and eventually eliminate) contributions from the modular Hamiltonian to relative entropy. In this case, the entanglement entropy inherits the monotonicity and positivity properties of relative entropy, and this can be used to understand RG flows. We will apply this idea to flows of the type (\ref{eq:RG1}). In this section we study the dependence of relative entropy on the Cauchy surface, and analyze in detail the null limit. In Sec. \ref{sec:consequences} we will consider the consequences for the RG.

%%%%%%%%%%%%%%%%%%%%%%%%%%%%%%%%%%%%%%%%%%%%%%%%%%%%%%%%%%
%%%%%%%%%%%%%%%%%%%%%%%%%%%%%%%%%%%%%%%%%%%%%%%%%%%%%%%%%%
\subsection{Reduction to a spatial region of two states of different theories}\label{subsec:reduction}

In order to clarify the dependence of relative entropy on time, let us first consider only one QFT and review the standard way the state reduction is achieved in space-time. We can describe the operator content of the theory ${\cal T}_0$ in any global Cauchy surface $\Sigma_{gl}$ (where $gl$ stands for global) by a set of fields we call generically $\phi_\lambda(x)$, with $x\in \Sigma_{gl}$, that form a complete set of generators for the operators in the Hilbert space. These set of operators may include time derivatives of the fields, or to adapt this description to $\Sigma_{gl}$, derivatives in the normal direction to $\Sigma_{gl}$ instead of time derivatives. For any  $\Sigma\subseteq \Sigma_{gl}$ we can form the algebra ${\cal A}_\Sigma$ generated by polynomials of the operators localized in this spatial region. Given a global state $\rho^0_{gl}$, its restriction to ${\cal A}_\Sigma$ gives the reduced state $\rho^0_\Sigma$ to $\Sigma$. This is just the state\footnote{We are using the abstract definition of a state as a positive normalized linear function on the operators of an algebra with values in the complex numbers. See for example \cite{petz2007quantum}. This is a density matrix once a basis has been selected to write the operators. We often interchange between the abstract state and its the density matrix representation.} 
on ${\cal A}_\Sigma$ that gives place to the same expectation values than the global state would give for all operators in this region. Notice that we can take an arbitrary state and have not used the dynamics or the Hamiltonian of the theory in this construction.

\begin{figure}[t]
\begin{center}  
\includegraphics[scale=0.60]{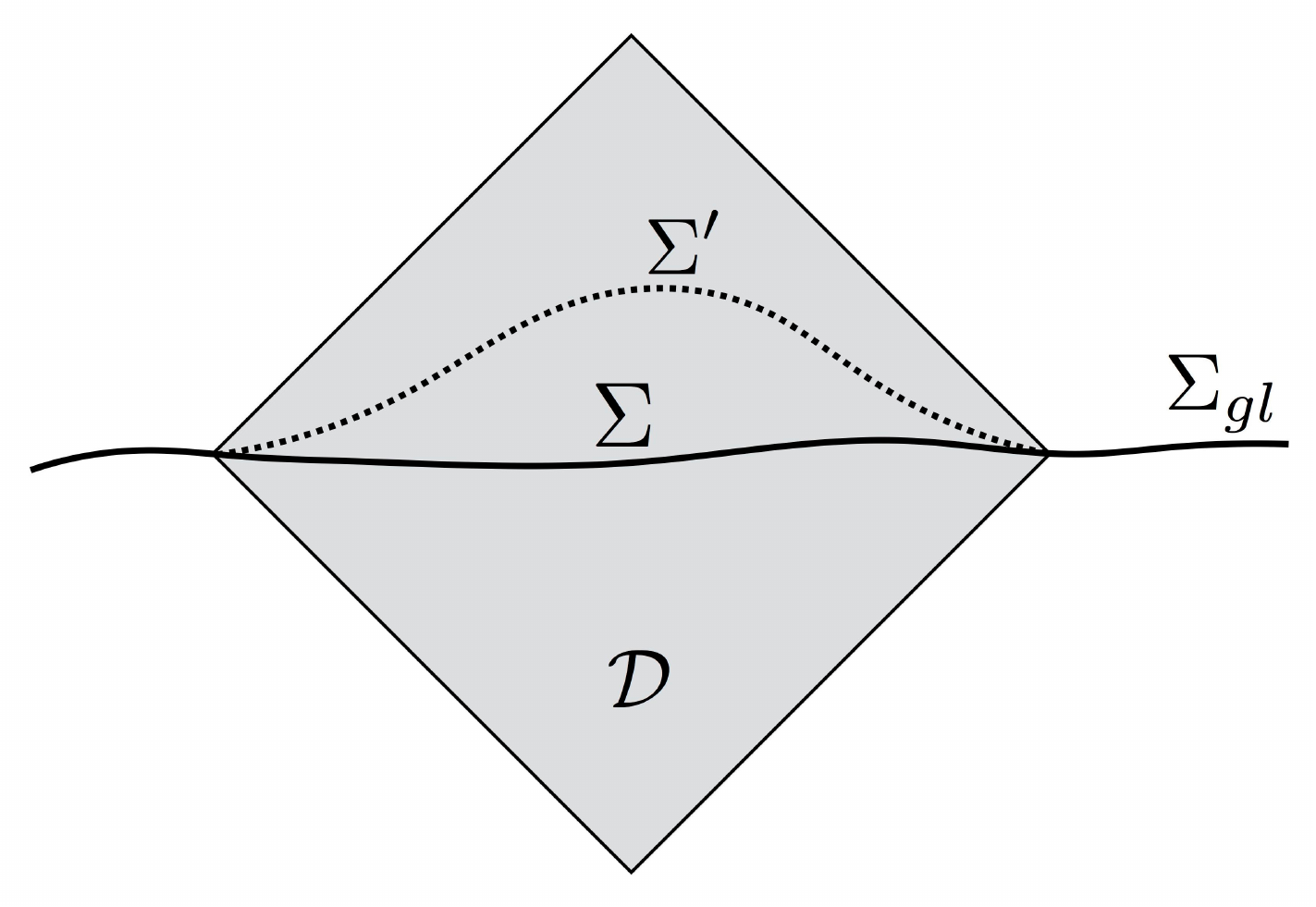}
\captionsetup{width=0.9\textwidth}
\caption{Different Cauchy surfaces $\Sigma$ and $\Sigma'$ with the same causal domain of dependence $\mc D$; $\Sigma_{gl}$ is a global Cauchy surface.}
\label{fig:cauchy1}
\end{center}  
\end{figure}

Let us consider another spatial surface $\Sigma^\prime$ with the same causal development ${\cal D}$ as $\Sigma$ (see figure  \ref{fig:cauchy1}). In the Heisenberg representation, states do not depend on time and operators obey the usual Heisenberg equations of motion. Operators localized at points in $\Sigma^\prime$ belong to the causal development of $\Sigma$ and can be written in terms of the ones in $\Sigma$ using the equations of motion. This identification depends on the Hamiltonian of the theory. Taking this into account we see that the algebra generated by the fields on $\Sigma^\prime$ coincides with the one on $\Sigma$. Since the global state does not depend on any choice of Cauchy surface in the Heisenberg representation, and the algebra on the two surfaces is the same, we conclude the reduced states $\rho^0_\Sigma$ and $\rho^0_{\Sigma^\prime}$ are the same. That is, they give the same expectation values, for the same operators on the same algebra, where operators are identified between $\Sigma$ and $\Sigma^\prime$ using the equations of motion. Therefore, the entanglement entropies are the same, if they are regularized in the same manner (for instance, by using the mutual information to provide a geometric cutoff). Relative entropies for two different states in this theory will be independent on the choice of Cauchy surface. The subalgebra of operators, and the reduced states, can then be thought of as functions of the causal completion or causal development ${\cal D}$ of $\Sigma$ (which coincides with the one of $\Sigma^\prime$), rather than functions of Cauchy surfaces. 

Now, let us modify the Hamiltonian by adding a source term as in (\ref{eq:RG1}), in such a way that we can still describe a generating basis for the operators in a Cauchy surface by the same set of fields, that we call $\tilde{\phi}_\lambda(x)$ for this new theory ${\cal T}_1$. We might need to impose a cutoff to do so.\footnote{In general it is also necessary to impose a cutoff to define the algebras in a sharp time slice as we are doing here. As it is the usual practice in QFT, the question of the existence of the relative entropy in the continuum limit will be translated into the question about the finiteness of this quantity as we remove the cutoff.} 
Let us also consider the Heisenberg representation with respect to this new Hamiltonian, and another global state $\rho^1_{gl}$ for this new theory.
Again, $\rho^1_\Sigma$ and its entropy will be invariant in changing Cauchy surfaces $\Sigma$ and $\Sigma^\prime$ (the density matrix representing this state can of course change if we change basis). Accordingly we will drop the subindex $\Sigma$ of these states.

If we want to compare the two states of the two theories with relative entropy we need to identify the Hilbert spaces, or equivalently, the algebra of operators of the theories, in a precise way. For doing this identification we will use a Cauchy surface. Given a Cauchy surface $\Sigma$ we naturally identify the field operators $\phi_\lambda(x)$ with $\tilde{\phi}_\lambda(x)$ for $x\in \Sigma$. Formally, the identification $\phi_\lambda(x) \leftrightarrow  \tilde{\phi}_\lambda(x)$ is carried out by a unitary operator $U_{\Sigma}$ that maps Hilbert spaces and operators between theories such that
\be\label{ident} 
   U_{\Sigma} \tilde{\phi}_\lambda(x) U_{\Sigma}^\dagger= \phi_\lambda(x)\,\hspace{.8cm} x\in \Sigma.
\ee  
The expectation values of the operators $\phi_\lambda(x)$ on $\Sigma$ computed with the two states $\rho^0$ and $U_\Sigma\rho^1 U_\Sigma^\dagger$ define two different reduced states on the same algebra. The state $U_\Sigma\rho^1 U_\Sigma^\dagger$ gives just the same expectation values on the fields of the first theory as $\rho^1$ on the fields of the second theory,
\be
\textrm{tr}(U_\Sigma\rho^1U_\Sigma^\dagger \phi_{\lambda_1}(x_1)...\phi_{\lambda_m}(x_m))=\textrm{tr}(\rho^1 \tilde{\phi}_{\lambda_1}(x_1)...\tilde{\phi}_{\lambda_m}(x_m))\,, \hspace{.5cm} x_i \in \Sigma\,.
\ee
We can then compute the relative entropy $S(U_\Sigma\rho^1 U_\Sigma^\dagger|\rho^0)$. Analogously we can compute $S(\rho^1|U_\Sigma^\dagger\rho^0 U_\Sigma)$, with the same result. This follows from the invariance of relative entropy under the simultaneous change of the states by the same unitary. 

To be clear, both states, $\rho_0$ of ${\cal T}_0$ and  $\rho_1$ of ${\cal T}_1$ define expectation values for operators in ${\cal D}$ in each theory.  To compute the relative entropy between these states we map the algebras by identifying its local basis elements: $\phi_\lambda(x) \leftrightarrow  \tilde{\phi}_\lambda(x)$, that is, with (\ref{ident}). We can write this relative entropy simply as
\begin{equation} \label{def} 
S_\Sigma(\rho^1 |\rho^0) \equiv \left. S(\rho^1 |\rho^0)\right|_{\phi_\lambda(x) \leftrightarrow \tilde{\phi}_\lambda(x)\,;\ x\in \Sigma} 
\end{equation}

This construction does not differ form the usual way relative entropy is computed in lattice systems. For instance, we can imagine a lattice on the surface where spin degrees of freedom sit at the vertices. We have two states, coming for example from the fundamental states of two different Hamiltonians. Then we can compute the relative entropy between these two states by assuming the spin operators are identified.

We do this at each Cauchy surface under consideration. If we pick another Cauchy surface $\Sigma^\prime$ in the same Causal domain $\cal D$ of $\Sigma$, the relative entropy we have defined will depend on the Cauchy surface; $S_\Sigma(\rho^1 |\rho^0)$ will differ from $S_{\Sigma^\prime}(\rho^1 |\rho^0)$.
 The reason for this change is that the identification of local basis elements $\phi_\lambda(x^\prime) \leftrightarrow  \tilde{\phi}_\lambda(x^\prime)$, $x^\prime\in \Sigma^\prime$, will be different from the identification in $\Sigma$, or, in the above language, $U_\Sigma$ is different from $U_{\Sigma^\prime}$.  
 This is because the local fields $\phi_\lambda(x)$ of $\Sigma$ can be expressed, by the equation of motion of ${\cal T}_0$, as a certain non-local function $\phi_\lambda(x)=F^{\Sigma^\prime}_0[\phi(x^\prime)]$ of the fields with $x^\prime\in\Sigma^\prime$, but 
 for the theory ${\cal T}_1$ we have a different function $\tilde{\phi}_\lambda(x)=F^{\Sigma^\prime}_1[\tilde{\phi}(x^\prime)]$ to express the fields, because the theories ${\cal T}_0$ and ${\cal T}_1$ have different equation of motions. 
Identifying $ \phi_\lambda(x)\leftrightarrow \tilde{\phi}_\lambda(x)$ on $\Sigma$ is to identify $F^{\Sigma^\prime}_0[\phi(x^\prime)]$ with $F^{\Sigma^\prime}_1[\tilde{\phi}(x^\prime)]$. Since $F_0^{\Sigma^\prime}$ and $F_1^{\Sigma^\prime}$ are different functions, this is not compatible with the identification of local fields on $\Sigma^\prime$. 
  As a result, identifying local operators in different surfaces leads to different relations between Hilbert spaces. 
 
 In a general interacting theory it is difficult to obtain $F^{\Sigma^\prime}$ explicitly. Fortunately we will not need it. As an example where the evolution between surfaces can be made explicit, consider as ${\cal T}_0$  a free scalar field of mass $m_0$. We have
\be\label{ex} 
\phi(x)= i\int_{\Sigma^\prime} d^{d-1}x^\prime\,\sqrt{h} \, \left( \eta^\mu \partial^x_\mu C_0(x-x^\prime) \phi(x^\prime)+ C_0(x-x^\prime)\eta^\mu \partial_\mu \phi(x^\prime)\right)\,, 
\ee
where $x \in {\cal D}$ and $x^\prime \in \Sigma^\prime$, $h$ is the induced metric on $\Sigma^\prime$, $\eta^\mu$ is the unit vector normal to $\Sigma^\prime$, and 
\be
C_0(x-x^\prime)=[\phi(x),\phi(x^\prime)]
\ee
is the commutator function of the scalar field of mass $m_0$. The normal derivative $\eta^\mu \partial_\mu \phi(x^\prime)\equiv \pi(x^\prime)$ is the momentum operator adapted to $\Sigma^\prime$, and has to be consider an independent operator on this surface ( $\lbrace \phi_\lambda(x^\prime)\rbrace_{\lambda=1,2} = \lbrace \phi(x^\prime), \pi(x^\prime) \rbrace $). We can consider as the theory ${\cal T}_1$ a scalar field with a different mass $m_1$. This has a different commutator function $C_1$ in place of $C_0$ in (\ref{ex}), giving $\tilde{\phi}(x)$ as a different combination of fields in $\Sigma^\prime$.

%%%%%%%%%%%%%%%%%%%%%%%%%%%%%%%%%%%%%%%%%%%%%%%%%%%%%%%%%%
%%%%%%%%%%%%%%%%%%%%%%%%%%%%%%%%%%%%%%%%%%%%%%%%%%%%%%%%%%
\subsection{Conformal interaction picture}\label{subsec:int}

The previous construction based on the Heisenberg representation makes manifest the dependence of the relative entropy on the choice of Cauchy surface. However, it is not the most convenient approach for concrete calculations. For this reason, we now present an equivalent discussion in terms of a ``conformal interaction picture'', which is a generalization of the standard interaction picture representation of QFT.

In the interaction picture of weakly coupled QFT, the Hamiltonian is split into a free part $H_0$ and an interacting part $H_{int}$. Operators in the interaction basis are chosen in the Heisenberg representation of the free Hamiltonian $H_0$, and states then evolve unitarily according to the evolution operator for $H_{int}$,
\be\label{eq:U1}
U(t_1, t_2)= T\left(\exp\lbrace-i \int_{t_1}^{t_2} dt\,H_{int}(t)\rbrace \right)\,.
\ee
Here $T$ denotes time-ordering, and $H_{int}$ is written in the interaction picture. This leads to the standard perturbative expansion around the free theory.

In our case, instead of a free theory we have a CFT, and the interaction is given by the perturbation $H_{int}= \int d^{d-1} x\,g\,\mathcal O(x)$. We then define a conformal interaction picture where operators are in the Heisenberg representation of the CFT Hamiltonian, while the state evolution is given by $H_{int}$. In more detail, let us denote the Heisenberg vacuum of the CFT ${\cal T}_0$ by $|0\rangle$ and its Heisenberg operators by $\phi_\lambda(x)$ as in the previous section. For the perturbed theory ${\cal T}_1$, we note the corresponding objects by $|\Omega\rangle$ and $\t \phi_\lambda(x)$. Time-ordered correlators of ${\cal T}_1$ become, in the interaction picture,
\be\label{eq:corr-int}
\langle \Omega | T\lbrace \t \phi_1(x_1) \ldots \t \phi_n(x_n) \rbrace|\Omega\rangle = \frac{\langle 0| T \lbrace \phi_1(x_1) \ldots \phi_n(x_n) \,\exp\left[-i \int_{-\infty}^\infty dt H_{int}(t)\right]\rbrace|0\rangle}{\langle 0| T \lbrace \exp\left[-i \int_{-\infty}^\infty dt H_{int}(t)\right]\rbrace|0\rangle}\,.
\ee
The factor in the denominator\footnote{The integrals in time have to be done in a slightly imaginary direction to project onto the vacuum state.}
  arises from the evolution that maps $|0\rangle \langle 0|$ into $|\Omega\rangle \langle \Omega|$. In this way, an expectation value in ${\cal T}_1$ is reduced to the calculation of a correlation function in the CFT ${\cal T}_0$. In particular, for small $g$ the right hand side in (\ref{eq:corr-int}) can be evaluated using the standard rules of conformal perturbation theory.

We can now redo the steps in section \ref{subsec:reduction} in the interaction picture. The operators for ${\cal T}_0$ and  ${\cal T}_1$ are now the same, $\phi_\lambda(x)$, corresponding to the Heisenberg CFT operators. Therefore, and recalling the map (\ref{eq:corr-int}), we can now think in terms of two different states $\rho^0$ and $\rho^1$ in the same theory. For concreteness, consider reduced states on a spatial region associated to the vacuum states (it is easy to extend the following discussion to more general states). As before, we choose a
global Cauchy surface $\Sigma_{gl}$, and let $\Sigma$ be a part of it.
The Heisenberg vacuum of ${\cal T}_0$ gives a state $\rho^0$ that is independent of $\Sigma$. However, the state $\rho^1$ for ${\cal T}_1$ evolves explicitly with time.

For a surface of constant time, the evolution is given by (\ref{eq:U1}). For instance, the state at $t=0$ is given by
\be
\rho^1 = K\,\tr_{\bar \Sigma}\,U(0, -\infty) \vert 0\rangle \langle 0 \vert U(\infty,0)\,,
\ee
with $K$ a normalization factor that sets $\tr \rho^1=1$. For a more general surface, we can evolve the state using a source $g(x;\Sigma_{gl})$ that is nonzero and equals $g$ only for $x$ in the region of spacetime below the surface $\Sigma_{gl}$:
\be
\rho^1_\Sigma= K\,\tr_{\bar \Sigma}\, U(\Sigma_{gl}) \vert 0\rangle \langle 0 \vert U^\dag(\Sigma_{gl})\,,
\ee
and
\be
U(\Sigma_{gl})= T\left(\exp\lbrace-i \int d^dx\,g(x; \Sigma_{gl}) \mathcal O(x)\rbrace \right)\,.
\ee
From here we have for two surfaces, $\Sigma$ and $\Sigma^\prime$ to the future of $\Sigma$, 
\be
\rho^1_{\Sigma^\prime}= U_{\Sigma,\Sigma^\prime}\rho^1_\Sigma U_{\Sigma,\Sigma^\prime}^\dagger\,,
\ee
with
\be
U_{\Sigma,\Sigma^\prime}=T\left(\exp\lbrace-i \int_{V_{\Sigma\Sigma^\prime}} d^dx\,g\, \mathcal O(x)\rbrace \right)\,,
\ee
where $V_{\Sigma\Sigma^\prime}$ is the spacetime region between $\Sigma$ and $\Sigma^\prime$.

This exhibits how the state $\rho^1_\Sigma$ depends explicitly on $\Sigma$ in the interaction picture; expectation values calculated with this state (such as the relative entropy) will also depend on the Cauchy surface.
 
%%%%%%%%%%%%%%%%%%%%%%%%%%%%%%%%%%%%%%%%%%%%%%%%%%%%%%%%%%
%%%%%%%%%%%%%%%%%%%%%%%%%%%%%%%%%%%%%%%%%%%%%%%%%%%%%%%%%% 
\subsection{Modular Hamiltonian}\label{subsec:modularH}

It is convenient to express the relative entropy by the equivalent expression
\begin{equation}\label{DSmasDH} 
S(\rho^1|\rho^0) = \Delta \langle {\cal H}\rangle- \Delta S  \,,
\end{equation} 
where
 \begin{equation}\label{DS} 
 \Delta S = S(\rho^1) - S(\rho^0) \ ,    
 \end{equation}
 is the difference of von Newmann entropies, and
\begin{equation}\label{DH} 
\Delta \langle {\cal H} \rangle = \text{Tr}(\rho^1 {\cal H} ) - \text{Tr}(\rho^0 {\cal H} ) \ , 
\end{equation}
is the difference of the expectation values of the modular Hamiltonian 
\be
{\cal H} = -\log \rho^0 \,.
\ee
In (\ref{DS}) and (\ref{DH}) the states appear in the same order as they enter in the arguments of $S(\rho^1|\rho^0)$. 

In the present case, $\Delta S$ gives the difference between the entanglement entropies of the two vacuum states in the same region. This term does not depend on the choice of Cauchy surface. The dependence on $\Sigma$ comes exclusively from the expectation value of the modular Hamiltonian,\footnote{The following formulas are written in the conformal interaction picture.}
\begin{equation}
\Delta \langle{\cal H}\rangle_\Sigma =\text{Tr} \left(  (\rho^1_\Sigma - \rho^0) {\cal H}\right)\ .
\end{equation}
${\cal H}$ is an operator in the theory ${\cal T}_0$. Its expectation value in the state $\rho^0$ is independent of the Cauchy surface. However, its expectation value using the second state $\rho^1_\Sigma$ depends on which surface we have identified operators.

In order to proceed we will choose ${\cal T}_0$ to be a CFT, $\rho^0$ is its vacuum state, and restrict attention to the case where the boundary of $\Sigma$ is a $d-2$ dimensional sphere. The modular Hamiltonian for this case has a simple expression in terms of the energy momentum tensor $T_{\mu\nu}$ of the theory ${\cal T}_0$ \cite{Hislop:1981uh, Casini:2011kv},
\begin{equation}
\label{Hmod}
{\cal H} =  \int_\Sigma d\sigma\ \eta^\mu \xi^\nu T_{\mu\nu} \,. 
\end{equation}
Here $\eta^\mu$ is a norm one, future pointing, normal vector to the Cauchy surface $\Sigma$,  and $\xi^\nu$ is the conformal Killing vector corresponding to conformal transformations keeping the sphere fixed. For a sphere centered at the origin in the plane $x^0=0$, its $(x^0, x^i)$ components are
\begin{equation}
\label{modflow}
\xi^\nu = \frac{\pi}{R}\left(  R^2 -(x^0)^2- (\vec{x})^2, -2 x^0 x^i \right) \ ,
\end{equation} 
where $R$ is the radius of the sphere. One can check that the current $j_\mu = \xi^\nu T_{\mu\nu}$ is conserved using that $T_{\mu\nu}$ is symmetric, conserved, and has zero trace. This makes ${\cal H}$ a conserved charge independent of the Cauchy surface in ${\cal T}_0$, but this is not the case when we evaluate its expectation value using $\rho^1_\Sigma$.  

In order to evaluate $\Delta \langle{\cal H}\rangle_\Sigma$ we need to understand the change in expectation value of the stress tensor $\Delta\langle T_{\mu\nu}(x)\rangle_\Sigma$. This is a local operator and its expectation value in the new state $\rho^1_\Sigma$ depends on the structure of the state (the correlation functions) near the point $x$ on this surface.  Then, we expect a local expression, that can involve only local tensors. These are $g_{\mu\nu}$ and all local geometrical quantities that can be constructed with the Cauchy surface, such as the normal $\eta^\mu$, the extrinsic and intrinsic curvatures, etc.. Given the Lorentz invariance of each vacuum in its respective theory, no other tensors can appear. 

However, curvature terms can only appear as corrections accompanied by positive powers of the cutoff, for example in the form $K_{ij}^2 \epsilon^2$, with $K_{ij}$ the extrinsic curvature of $\Sigma$ and $\epsilon$ a short distance cutoff. This is because we are evaluating the expectation value of a local operator for a QFT in flat space, and the shape of $\Sigma$ only enters in the correlation functions through the distance between points. For example, in a lattice regularization $T_{\mu\nu}$ can be written in terms of operators at a point and few of its neighbors, and the expectation value in the state $\rho^1_\Sigma$ depends only on short distance correlations functions on the lattice. We show some explicit examples for free fields in the Appendix. The curvature then only enters modifying the distance of nearby points, and is always accompanied by the cutoff. These terms can be neglected if the curvature is much smaller that the cutoff scale. We will always assume that this is the case. This is also necessary since we can define the position of the Cauchy surface only at scales larger than the cutoff.  

Therefore we have the general form
\be
\Delta\langle T_{\mu\nu}(x)\rangle_\Sigma =  k \left( \eta_\mu(x) \eta_\nu(x)-\frac{g_{\mu\nu}}{d}     \right) +{\cal O}(K^2 \epsilon^2)+...\,.\label{mudu}
\ee
We have used the fact that the stress tensor of the CFT is traceless. This expectation value depends on the Cauchy surface through the normal vector $\eta_\mu$, and this is crucial in order to have a traceless symmetric tensor in an otherwise Lorentz invariant computation. Note $\Delta\langle T_{\mu\nu}(x)\rangle_\Sigma$ does not transform as a Lorentz tensor unless $\Sigma$ is also transformed.
Eq.(\ref{mudu}) will be quite important for our arguments below. For this reason, in the Appendix we perform explicit calculations of $\Delta\langle T_{\mu\nu}\rangle$ for mass flows in free scalar field theories, and exhibit the dependence on the Cauchy surface.

Let us find out the possible behavior of the constant $k$ with the cutoff. If $k$ is divergent with the cutoff we expect a perturbative calculation would give its leading behavior. The reason is that the coupling $g$ in 
(\ref{eq:RG1}), responsible for deforming ${\cal T}_0$ into ${\cal T}_1$, is relevant and hence goes to zero in the UV.
Perturbative corrections start at second order in $g$ since $\langle T_{\mu\nu}\mathcal O\rangle=0$ for a primary operator in a CFT. Taking into account that the dimension $[k]=d$ and $[g]=d-\Delta$ , we have by dimensional analysis
\be
k\sim g^2 \epsilon^{d-2 \Delta}\,.\label{k}
\ee 
Therefore, the expectation value of the modular Hamiltonian between the CFT and the perturbed theory in a spatial surface is UV divergent for $\Delta\ge d/2$. For $\Delta< d/2$ we cannot compute $k$ perturbatively, but we expect a finite $k$.  

\begin{figure}[t]
\begin{center}
\includegraphics[scale=0.3]{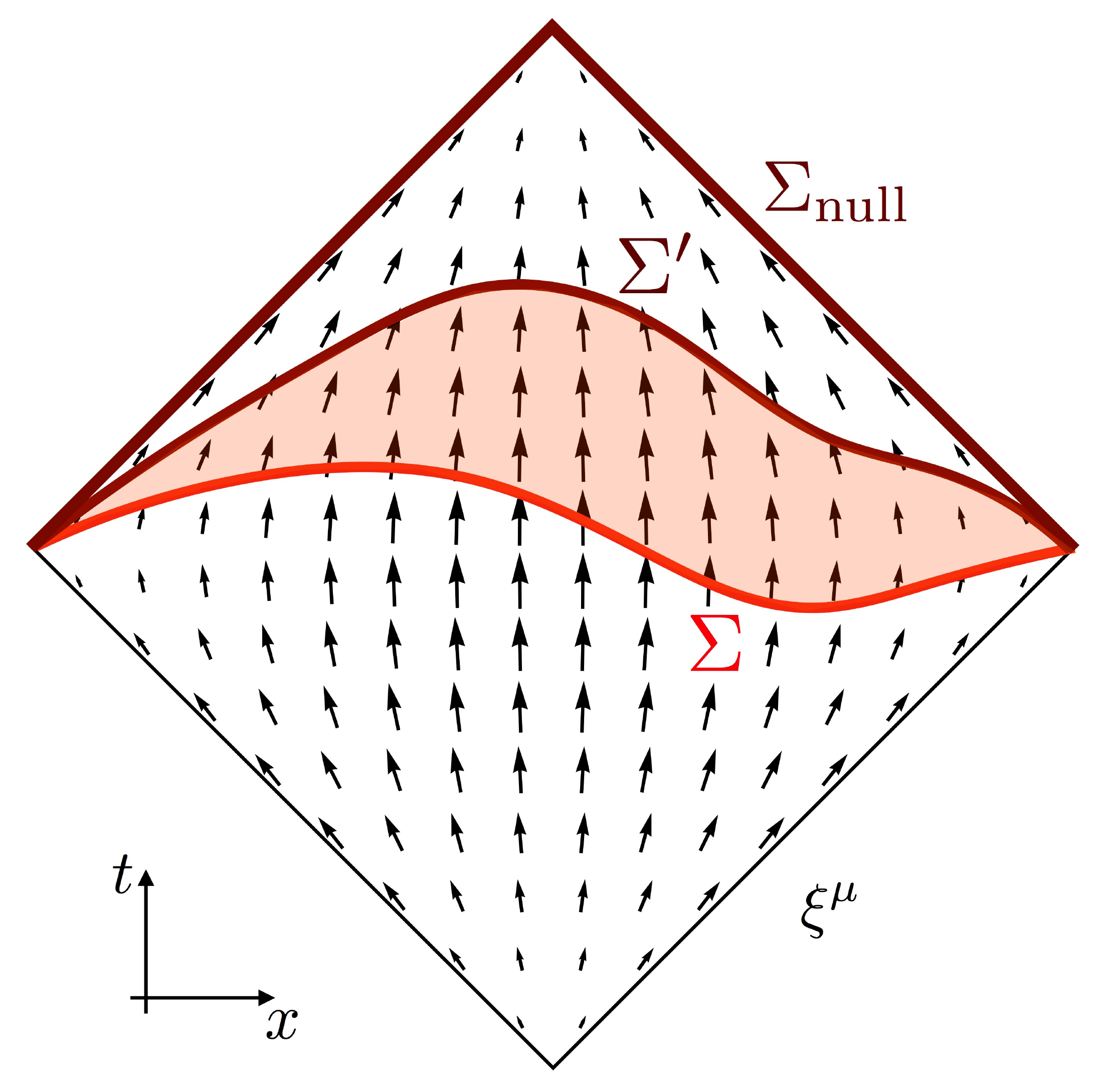} 
\caption{Modular flow vector field $\xi^\mu$ in the $d=2$ causal diamond of a spatial sphere. $\Delta \langle{\cal H}\rangle_{\Sigma}$ is described by the flux of $\xi^\mu$ through the Cauchy surface $\Sigma$. $\Delta \langle{\cal H}\rangle_{\Sigma_{\text{null}}} = 0$ on the null Cauchy surface $\Sigma_{\text{null}}$ for the range of perturbations discussed in the text. The divergence of $\xi^\mu$ integrated over the shaded region gives $\Delta {\cal H}_{\Sigma^\prime}-\Delta {\cal H}_\Sigma$ and hence the variation of the relative entropy with the surface.}
\label{fi2}
\end{center}
\end{figure}

The dependence of $\Delta \langle {\cal H} \rangle_\Sigma$ with the Cauchy surface $\Sigma$ follows from inserting (\ref{mudu}) into (\ref{Hmod})
\begin{equation}\label{DT} 
\Delta\langle {\cal H}\rangle_\Sigma =  \frac{k(d-1)}{d} \int_\Sigma d\sigma\, \eta^\mu \xi_\mu  \,. 
\end{equation}
Then, this simple geometrical dependence is described by the flux of the field $\xi_\nu$ through $\Sigma$ (see figure \ref{fi2}).
This changes because the flux in (\ref{DT}) is not constant and as a consequence of Gauss' theorem
\begin{equation}\label{gauss} 
\Delta \langle{\cal H}\rangle_{\Sigma^\prime}-\Delta \langle{\cal H}\rangle_\Sigma=\frac{k(d-1)}{d}\left(\int_{\Sigma^\prime} d\sigma\ \eta^\mu \xi_\mu - \int_\Sigma d\sigma\ \eta^\mu \xi_\mu\right) =  \frac{k(d-1)}{d} \int_{V_{\Sigma\Sigma^\prime}} dv\ (\partial \cdot \xi)\,,
\end{equation}
where
\be
(\partial\cdot \xi) = - \frac{2 \pi d}{R} x^0\,,
\ee
and $V_{\Sigma\Sigma^\prime}$ is the space-time region between the two surfaces.

The infrared behavior of the expectation value of the modular Hamiltonian follows from this integral. For the planar Cauchy surface at $x^0=0$ we get
\be\label{eq:Rd}
\Delta \langle{\cal H}\rangle=\frac{2\pi \tilde{\Omega}\, k}{d(d+1)} \,R^d\, ,
\ee 
where $R$ is the radius of the spherical entangling surface and 
\begin{equation}\label{areaS} 
\tilde{\Omega} = \frac{2 \pi^{\frac{d-1}{2}} }{ \Gamma(\frac{d-1}{2})}
\end{equation}
is the area of the unit sphere immersed in $\mathbb{R}^{d-1}$ ($S^{d-2}$ sphere).
The same super-extensive behavior $\sim R^d$ holds for other spatial surfaces that do not approach much to the null horizon of the causal development of the sphere.

%%%%%%%%%%%%%%%%%%%%%%%%%%%%%%%%%%%%%%%%%%%%%%%%%%%%%%%%%%
%%%%%%%%%%%%%%%%%%%%%%%%%%%%%%%%%%%%%%%%%%%%%%%%%%%%%%%%%% 
\subsection{The null limit}\label{subsec:null}

Having understood the general dependence on the Cauchy surface, we are now ready to approach the null limit. 
From the expression (\ref{modflow})  for $\xi^\nu$  and the definition of $\eta^\mu$, we find that on the limit of the null Cauchy surface for the sphere 
\begin{equation}
\left. (\eta^\mu  \xi_\mu )\right|_{\Sigma_{\text{null}}}=  0\,.
\end{equation}  
In fact, both vectors becomes null vectors on the null Cauchy surface. With this, and (\ref{DT}), we obtain the interesting result
\begin{equation}\label{H0} 
\Delta \langle {\cal H} \rangle_{\Sigma_{\text{null}}} = 0\ .
\end{equation}

This limit, however, is not necessarily justified because the coefficient in (\ref{DT}) can be divergent. As we mentioned above, we need to assume that the typical scale of curvature of $\Sigma$ is large with respect to the cutoff $\epsilon$. As we go to the null surface, the extrinsic and intrinsic curvature of a spatial surface will typically diverge. For example, a hyperboloid $(x^0)^2-(\vec{x})^2=a^2$ has a curvature scale of order $a^{-1}$, and the null limit is $a\rightarrow 0$. Put differently, we need that the cutoff scale $\epsilon$ is always much smaller than the total length across the surface $\Sigma$, in order for example, to associate the cutoff to a physical lattice on the surface. Hence, we need to keep $\epsilon\lesssim a$ as we take the null limit $a\rightarrow 0$. We can take the ratio $a/\epsilon$ to be some arbitrarily large number, but keep it fixed as we take the simultaneous limit $\epsilon\sim a \rightarrow 0$. This automatically keeps the curvature terms in (\ref{mudu}) under control. Given this, we should understand next when $\Delta \langle \mathcal H\rangle_\Sigma$ vanishes.

Let us examine the expression (\ref{DT}) in the null limit. For simplicity we consider as Cauchy surfaces a family of hyperboloids $\Sigma_a$ parametrized with the radius $a$, 
\begin{equation}\label{locus} 
\left( x^0 - \sqrt{a^2 + R^2}  \right)^2 - \vec{x}^2 = a^2\,, \ \ \ \ \ \ \ 0 < |\vec{x}| <R\,,\,\,\,\, x^0>0.
\end{equation}
The integral
\begin{equation}
\int_{\Sigma_a} d\sigma\, \eta^\mu \xi_\mu = \left\lbrace  \begin{array}{ll}
4 \pi a^2 \log(R/a)  + {\cal O}(a^3) & d=2    \\
\frac{2 \pi \tilde{\Omega} }{(d-1)(d-2)} a^2 R^{d-2} + {\cal O}(a^3) & d>2
\end{array}  \right. \ ,
\end{equation}
generically goes as $\sim a^2 R^{d-2}$ for small $a$. Plugging this back into (\ref{DT}) and using (\ref{k}) for $k$, we have 
\be
\Delta \langle {\cal H}\rangle_\Sigma\sim g^2 R^{d-2} \epsilon^{d-2 \Delta} a^{2}\,. \label{ll}
\ee
We see that using hyperboloids of constant $a$, the contribution of the modular Hamiltonian increases like the area $\sim R^{d-2}$ instead of the $R^d$ dependence in the surface $x^0=0$, (\ref{eq:Rd}). Eq. (\ref{ll}) gives a vanishing limit for $a\sim \epsilon\rightarrow 0$ provided that
\be\label{eq:window1}
\Delta<\frac{d+2}{2}\,.
\ee
That is, the null limit enlarges the window where the modular Hamiltonian gives a finite contribution from $\Delta< d/2$ to $\Delta <(d+2)/2$. In this window in fact this contribution vanishes in the null limit. We do not have control of the null limit for $\Delta\ge (d+2)/2$.   

In some special theories having a UV fixed point with free scalars, the modular Hamiltonian has an additional boundary term \cite{Lee:2014zaa, Lewkowycz:2014jia, Herzog:2014fra, Casini:2014yca}. This term scales like the area $R^{d-2}$ and does not depend on the Cauchy surface. Then it does not vanish in the null limit. However, this does not alter the conclusions about the relative entropy we want to make in this section. We discuss boundary terms in the modular Hamiltonian in more detail in the Appendix.   

%%%%%%%%%%%%%%%%%%%%%%%%%%%%%%%%%%%%%%%%%%%%%%%%%%%%%%%%%%
%%%%%%%%%%%%%%%%%%%%%%%%%%%%%%%%%%%%%%%%%%%%%%%%%%%%%%%%%%
\subsection{Entanglement entropy and regimes of relative entropy}\label{subsec:regimes}

Let us now briefly analyze the contribution of the entanglement entropy to the relative entropy. As we mentioned before, this does not depend on the Cauchy surface. The contribution of the entanglement entropy, in contrast to the one of the modular Hamiltonian, will generically be a complicated function of $R$ that depends on the full RG running of the model. We will say more about the entanglement entropy in the next section; however, the main features are well known. At the fixed points its leading term is proportional to the area, except for $d=2$ where it can grow logarithmically with $R$. We can ask when the EE will give a finite or divergent contribution. Again we expect that in the divergent case we can do a perturbative treatment. The divergent terms are going to be proportional to the boundary area since divergences are related to local entanglement that is extensive on the boundary of the region. Then we expect on dimensional grounds
\be\label{eq:DeltaS1}
\Delta S\sim g^2 R^{d-2} \epsilon^{d+2-2\Delta}\,.
\ee  
The allowed window for having finite $\Delta S$ is $\Delta< (d+2)/2$. This is well known from holographic calculations \cite{Hung:2011ta, Lewkowycz:2012qr, Liu:2013una} and direct computations of the renormalization of the area terms \cite{Hertzberg:2010uv, Rosenhaus:2014ula, Rosenhaus:2014zza, Casini:2014yca, Casini:2015ffa}. This coincides with the window (\ref{eq:window1}) for having vanishing $\Delta \langle {\cal H}\rangle$ in the null limit. We do not know of a deeper reason for this agreement. 

With this information and the one of the modular Hamiltonian we can summarize the different regimes for relative entropy between the two theories. 

First, for spatial surfaces (flat, or with curvature $\sim R^{-1}$) the relative entropy is dominated by the contribution of the modular Hamiltonian for large distances. In the infrared it grows superextensively as $R^d$. It is UV finite only for the window of perturbations with dimensions $\Delta< d/2$. For this range of $\Delta$ and at short distances, the entanglement entropy is finite; conformal perturbation theory then gives $\Delta S \sim g^2 R^{2(d-\Delta)}$, which goes to zero faster than $R^d$ for small $R$. The modular Hamiltonian thus dominates over the entanglement entropy at all scales for $\Delta<d/2$. Since the entanglement entropy is independent of $\Sigma$, the relative entropy changes with Cauchy surface in a simple geometric form as the modular Hamiltonian,
\begin{equation}\label{gauss1} 
S(\rho^1|\rho^0)_{\Sigma^\prime}-S(\rho^1|\rho^0)_\Sigma=  -\frac{2 \pi k(d-1)}{R} \int_{V_{\Sigma\Sigma^\prime}} dv\ x^0\,.
\end{equation}

On the other hand, the limit of relative entropy on null surfaces is finite for dimensions $\Delta< (d+2)/2$, extending the range $\Delta<d/2$ of spatial surfaces. In this window the contribution of the modular Hamiltonian vanishes and the relative entropy is entirely due to the entanglement entropies $S(\rho^1|\rho^0)=-\Delta S$. It grows as the area $\sim R^{d-2}$ in the infrared. The null relative entropy is finite for the same window in which it can be defined as a limit from the relative entropy of spatial surfaces, $\Delta<(d+2)/2$.

The result $S(\rho^1|\rho^0)_{\Sigma_{\text{null}}}=-\Delta S$ (or $\Delta \langle {\cal H} \rangle_{\Sigma_{\text{null}}} = 0$) gives to the null surface a special status. The relative entropy computed on it do not distinguish the vacuum states $\rho_1$, $\rho_0$ as much as when computed in other (spatial) surfaces of the same causal domain. The reason for this is that, as we take the null limit, correlations in the direction that is getting null become short distance correlations, and then are less efficient in distinguishing the state from its UV limit.

%%%%%%%%%%%%%%%%%%%%%%%%%%%%%%%%%%%%%%%%%%%%%%%%%%%%
%%%%%%%%%%%%%%%%%%%%%%%%%%%%%%%%%%%%%%%%%%%%%%%%%%%%
%%%%%%%%%%%%%%%%%%%%%%%%%%%%%%%%%%%%%%%%%%%%%%%%%%%%
\section{Consequences for the entanglement entropy}\label{sec:consequences}

The previous result $\Delta \langle {\cal H} \rangle_{\Sigma_{\text{null}}} = 0$ in the window
\be\label{eq:window2}
\frac{d-2}{2}\le \Delta<\frac{d+2}{2}
\ee
implies that
\begin{equation}\label{SrelDS} 
S(\rho^1|\rho^0) = -\Delta S = S(\rho^0) - S(\rho^1)
\end{equation}
on a null Cauchy surface. This reveals that $-\Delta S$  has the positivity and monotonicity properties of the relative entropy,
\be\label{eq:resultS}
-\Delta S \ge 0\;,\;\frac{d \, \Delta S}{d R} \le 0\,.
\ee
In this section we explore the consequences of this result in two and higher dimensions. For $d=2$ we find a simple alternative proof of the $c$-theorem, while for $d>2$ this will lead to the monotonicity of the area term in the entanglement entropy.

%\begin{eqnarray}
%&-\Delta S =  S(\rho_0) - S(\rho_1) \geqslant  0 \ ,& \label{posit}  \\
%& \frac{d \, \Delta S}{d R}  \leqslant  0 \ .\label{monot} 
%\end{eqnarray}
%From this it follows a simple alternative proof of the $c$-theorem in $d=2$ space-time dimensions and, for $d>2$, %the proof that the coefficient of the area term in the entanglement entropy decreases along the RG flow between %fixed points.

%%%%%%%%%%%%%%%%%%%%%%%%%%%%%%%%%%%%%%%%%%%%%%%%%%%%%%%
%%%%%%%%%%%%%%%%%%%%%%%%%%%%%%%%%%%%%%%%%%%%%%%%%%%%%%%
\subsection{A simple proof of the $c$-theorem}\label{subsec:cthm}

Let us consider the implications of (\ref{SrelDS}) for RG flows in $d=2$ spacetime dimensions. In this case, the window (\ref{eq:window2}) becomes $0<\Delta<2$, capturing all possible deformations by relevant operators. 

We take the theory ${\cal T}_0$ as an UV 2d CFT with central charge $c_{UV}$. We recall that, in this case, the entanglement entropy for an interval of size $R$ is of the form
\be\label{eq:Scft}
S(R) = \frac{c_{UV}}{3}\,\log(R/\epsilon)+c_0\,,
\ee
where $\epsilon$ is a short-distance cutoff and $c_0$ is a nonuniversal constant. 

In contrast, the entropy for ${\cal T}_1$ will have a more complicated radial dependence because it undergoes a nontrivial RG flow. However, at distances much longer than the typical mass scale $m\sim g^{-1/(d-\Delta)}$ of the RG flow, ${\cal T}_1$ goes to the IR fixed point of central charge $c_{IR}$. Taking into account that the UV divergences are still controlled by the UV fixed point of central charge $c_{UV}$, the EE for ${\cal T}_1$ at large distances is given by
\be\label{IR} 
S(R) = \frac{c_{IR}}{3}\,\log(m R)-\frac{c_{UV}}{3}\,\log(m\epsilon) +\textrm{const.}
\ee

Subtracting (\ref{eq:Scft}) to (\ref{IR}), we obtain the difference in EE between both theories at long distances is given by
\be
\Delta S\approx \frac{c_{IR}-c_{UV}}{3}\,\log(m R)\,,
\ee
up to terms that are subleading in $R$ (that we drop in what follows). From (\ref{eq:resultS}), $\Delta S<0$ and so we deduce that
\be
c_{IR} < c_{UV}\,.
\ee
This provides a new derivation of Zamolodchikov's $c$-theorem \cite{Zamolodchikov:1986gt} using the relative entropy on null surfaces.

%%%%%%%%%%%%%%%%%%%%%%%%%%%%%%%%%%%%%%%%%%%%%%%%%%%%%%%
%%%%%%%%%%%%%%%%%%%%%%%%%%%%%%%%%%%%%%%%%%%%%%%%%%%%%%%
\subsection{Monotonicity of the area term in entanglement entropy}\label{subsec:area}

Having understood the result for $d=2$, let us now consider QFTs in $d>2$. Note that for $d>2$ the restriction (\ref{eq:window2}) puts an upper bound $\Delta<(d+2)/2$ on the dimensions of RG perturbations. When $(d+2)/2<\Delta<d$, the perturbation is still relevant but the change in the modular Hamiltonian no longer vanishes; it is then not clear whether $-\Delta S$, which is also divergent in this range, inherits the monotonicity and positivity properties of the relative entropy. It would be interesting to study in more detail the regime $(d+2)/2<\Delta<d$, looking for possible cancellations of divergences, but in this work we restrict for simplicity to $\Delta<(d+2)/2$.

The EE for a QFT on a sphere of radius $R$, much bigger than all the length scales of the theory, is extensive on the boundary of the sphere, and hence
\be
S(R) = \mu\, R^{d-2}+ \ldots
\ee
where $\mu$ is a constant of mass dimension $d-2$, and `$\ldots$' are terms subleading in $R$. We want to understand properties of this area term along RG flows.

For a CFT such as theory ${\cal T}_0$ above, dimensional analysis dictates that
\be\label{eq:kT0}
\mu_{UV} = \frac{k_0}{\epsilon^{d-2}}\,,
\ee
where $k_0$ is a nonuniversal constant. On the other hand, theories with RG flows have additional mass scales that can also enter here. For ${\cal T}_1$ this is determined by $g$, the coefficient of the relevant perturbation. If conformal perturbation theory applies, the first correction is of order $g^2$, and hence we expect
\be
\mu_{IR} = \frac{k_0}{\epsilon^{d-2}}+g^2 \frac{k_1}{\epsilon^{2(\Delta-\frac{d+2}{2})}}+\ldots
\ee
See also (\ref{eq:DeltaS1}). The second term is divergent for $\Delta>(d+2)/2$, which is outside the range of dimensions (\ref{eq:window2}) under consideration. Instead, for $\Delta<(d+2)/2$, the contribution to the area term sourced by the RG will be finite, 
\be\label{eq:kT1}
\mu = \frac{k_0}{\epsilon^{d-2}}+k_1 m^{d-2}\,.
\ee
The dimensionless coefficient $k_1$ is in general non perturbative.

Comparing ${\cal T}_0$ and ${\cal T}_1$ through the relative entropy on a null surface implies $\Delta S <0$; this says that the coefficient of the area term decreases along RG flows, $\Delta \mu<0$, or
\be\label{eq:areathm}
\mu_{UV}> \mu_{IR}\,.
\ee
We call this the \textit{area theorem}. Note that the nonuniversal divergent term proportional to $1/\epsilon^{d-2}$ is the same in both theories, and hence it cancels out from this inequality. Therefore the finite renormalization in the area term in ${\cal T}_1$ has to be negative, $k_1 m^{d-2}<0$. We also note that the monotonicity condition $\frac{d \, \Delta S}{d R} \le 0$ does not give rise to new inequalities in this analysis of the IR behavior. For $d=2$ and $d=3$ eq. (\ref{eq:areathm}) also follows from strong subadditivity \cite{Casini:2004bw, Casini:2012ei}. 

This result has some interesting implications for gravity. The idea that part of the black hole entropy is due to entanglement entropy, suggests that the universal area term in the EE should agree with the renormalization of Newton's constant. This was made more precise in \cite{Rosenhaus:2014ula, Rosenhaus:2014zza, Casini:2014yca, Casini:2015ffa}, who related 
the Adler-Zee formula \cite{Adler:1982ri, Zee:1980sj},
\be\label{eq:AZ}
\Delta((4 G_N)^{-1})=-\frac{\pi}{d(d-1)(d-2)} \int d^d x\,x^2 \langle \Theta(x) \Theta(0) \rangle\,,
\ee
(where $\Theta(x) = T_\mu^\mu(x)$ is the trace of the stress tensor) to the finite part of the area term in the EE. These derivations use the first law of EE \cite{Blanco:2013joa} or holography \cite{Casini:2015ffa}.

From our approach, the universal part of the area term (given by $\Delta \mu=\mu_{IR}-\mu_{UV}$) is proven to be negative due to its relation to relative entropy. This does not use positivity of the stress-tensor two-point function, as in (\ref{eq:AZ}), and does not need to go through the first law of EE or holography. The situation is analogous to what happened in $d=2$, where positivity of the stress-tensor two-point function leads to the $c$-theorem \cite{Cappelli:1990yc}, while our proof relied on positivity of the relative entropy. In fact, the derivation based on the relative entropy emphasizes the common origin between the $c$-theorem and the area theorem, something that was also seen in the holographic context in \cite{Casini:2015ffa}. Furthermore, our approach identifies $\Delta \mu$ with a well-defined continuum quantity, and suggests further connections between quantum corrections to gravity and relative entropy.

%%%%%%%%%%%%%%%%%%%%%%%%%%%%%%%%%%%%%%%%%%%%%%%%%%%%%%%%%%%%%%%%%
%%%%%%%%%%%%%%%%%%%%%%%%%%%%%%%%%%%%%%%%%%%%%%%%%%%%%%%%%%%%%%%%%
%%%%%%%%%%%%%%%%%%%%%%%%%%%%%%%%%%%%%%%%%%%%%%%%%%%%%%%%%%%%%%%%%
%%%%%%%%%%%%%%%%%%%%%%%%%%%%%%%%%%%%%%%%%%%%%%%%%%%%%%%%%%%%%%%%%
\section{Conclusions}\label{sec:concl}

In this work we have shown that the $c$-theorem in $d=2$ and the decrease in the area term of the entanglement entropy between short and large distances are required by positivity and monotonicity of relative entropy. These results coincide with analogous results that use either reflection positivity of stress tensor correlators or strong subadditivity of entanglement entropy. However, as a bonus, the present proof relying on relative entropy gives a more direct QIT interpretation for the irreversibility of the RG: it corresponds to an increased distinguishability of vacuum states in a region as this region gets larger, allowing more operators to be used to distinguish states.  

In this sense, these monotonicity properties of the RG are a common quantum mechanical phenomenon. However, relativity and QFT enter crucially in the proof, in the fact that we needed to compare the states on null surfaces. Otherwise the  
 relative entropy distinguishes the states too much, giving non interesting information. The null surface decreased distinguishability in such a way that relative entropy turns out to be reduced to minus the difference in entanglement entropies. The reason the vacuum of the theory and the one of its CFT ultraviolet fixed point get more similar when compared on the null surface is physically clear. The correlators along null directions are UV correlators and cannot be used to distinguish them. Only correlation functions in the transverse directions matter.

This relative entropy in the null limit is finite only for $\Delta<(d+2)/2$. Otherwise correlators are different enough at arbitrarily short distances to allow for perfect distinguishability. When the relative entropy between the two vacuum states is not finite we may think they live in  ``different Hilbert spaces''.\footnote{A more precise mathematical statement would be that the local algebras of operators on the sphere are in two disjoint representations for the two theories.} For $\Delta<(d+2)/2$ this is not the case. However, for large regions relative entropy grows at least as $R^{d-2}$. Indeed, it is necessary to have divergent relative entropy for the full space, as in this limit we have two different pure states. It would be very interesting to develop techniques that could be applied to the full range of dimensions.

When the renomalization of the area term is finite, the result can be interpreted as an increase of Newton's constant towards the IR, due to QFT effects. This implies anti-screening of gravity.     
But at the same time it shows that the area term cannot be purely induced and finite, since it would be negative, and we would have a negative Newton constant. The entropy cannot be negative and needs an additional positive UV term to compensate for the sign, and the same should occur with the Newton constant. Of course this is an old problem (see \cite{Adler:1982ri} for example) and we just see it in a new perspective.  

It is interesting that the null relative entropy does not coincide with $-\Delta S$ for theories with free scalars in the UV, due to a boundary term in the modular Hamiltonian. In the Appendix we show calculations that suggest that taking the relative entropy as a form of regularized entropy restores the naive counting of divergent terms induced by the mass that fails for the free scalar.  In this sense the relative entropy gives a different regularization of entropy than, for example, mutual information.  However, the change with respect to other regularizations is a term exactly proportional to the area that, for example, does not alter the $c$-function. It corresponds to a specific choice of contact term in (\ref{eq:AZ}).

\section*{Acknowledgments}
This work was supported by CONICET PIP grant 11220110100533, Universidad Nacional de Cuyo, CNEA, and the Simons Foundation ``It from Qubit'' grant.

%%%%%%%%%%%%%%%%%%%%%%%%%%%%%%%%%%%%%%%%%%%%%%%%%%%%
%%%%%%%%%%%%%%%%%%%%%%%%%%%%%%%%%%%%%%%%%%%%%%%%%%%%
%%%%%%%%%%%%%%%%%%%%%%%%%%%%%%%%%%%%%%%%%%%%%%%%%%%%
\appendix

%%%%%%%%%%%%%%%%%%%%%%%%%%%%%%%%%%%%%%%%%%%%%%%%%%%%%%%
%%%%%%%%%%%%%%%%%%%%%%%%%%%%%%%%%%%%%%%%%%%%%%%%%%%%%%%
%%%%%%%%%%%%%%%%%%%%%%%%%%%%%%%%%%%%%%%%%%%%%%%%%%%%%%%
%%%%%%%%%%%%%%%%%%%%%%%%%%%%%%%%%%%%%%%%%%%%%%%%%%%%%%%  
\section{Free field examples}\label{sec:free}

In the main text we compared the two theories ${\cal T}_0$ and ${\cal T}_1$ in terms of the relative entropy. A crucial consequence of this analysis is the dependence on the choice of Cauchy surface, which enters via $\Delta \langle T_{\mu\nu} \rangle$ as in (\ref{mudu}). In this Appendix we illustrate how this happens in detail for free scalar fields. The required calculations can be performed explicitly, and we discuss the results with different cutoffs. We also show how the divergence in $\Delta S$ at $d=4$ is canceled by the boundary term in the modular Hamiltonian.

%%%%%%%%%%%%%%%%%%%%%%%%%%%%%%%%%%%%%%%%%%%%%%
%%%%%%%%%%%%%%%%%%%%%%%%%%%%%%%%%%%%%%%%%%%%%%
\subsection{Massless and massive scalar fields}\label{subsec:scalar}

In free field theory we can consider an RG flow given by perturbing a massless scalar with a mass term. The UV fixed point is simply the free massless scalar, and the relevant mass deformation triggers a flow that ends in a trivial gapped theory. In fact, it will be useful to consider a slightly more general setup, where ${\cal T}_0$ is the theory of a free scalar with squared mass $m_0^2$, while ${\cal T}_1$ is another theory with mass squared $m_1^2$. We want to compute the variation $\Delta \langle T_{\mu\nu} \rangle$ between both theories, with $T_{\mu\nu}$ the stress-tensor operator for ${\cal T}_0$.

Recall that a massive scalar field,
\be
S= \int d^dx \,\sqrt{g}\left( \frac{1}{2}g^{\mu\nu}\partial_\mu \phi\partial_\nu \phi - \frac{1}{2}m^2_0\phi^2\right)
\ee
has an energy-momentum tensor given by
\be\label{eq:Tmunu-scalar}
T_{\mu\nu}= \partial_\mu \phi \partial_\nu \phi - \frac{1}{2} g_{\mu\nu} \left((\partial \phi)^2-m^2_0 \phi^2\right)-\xi_c(\partial_\mu \partial_\nu-g_{\mu\nu}\partial^2)\phi^2\;\;,\;\;\xi_c=\frac{d-2}{4(d-1)}\,.
\ee
The last term is the improvement term. We have added it to have a traceless tensor in the massless limit.
We will compute $\Delta \langle T_{\mu\nu} \rangle$ with different regulators, and choose the spatial Cauchy surface $x_0=0$. 

A possible physical regulator is to use a point splitting associated to the choice of Cauchy surface; in the present case, we can split the points infinitesimally along the spatial surface. For this, we will need the Minkowskian propagator in  $d$ dimensions,
\be
  \langle  \phi(x)\phi(y) \rangle  = \langle T \phi(x)\phi(y) \rangle = \frac{1}{(2\pi)^{d/2}}\left( \frac{m}{|x-y|}\right)^{\frac{d}{2}-1} K_{\frac{d}{2}-1}(m |x-y|)
\ee
where the distance $|x-y|=\sqrt{(\vec{x}-\vec{y})^2 - (x_0-y_0)^2}$. The $T_{00}$ for a scalar field of mass $m_0$, with the point splitting regularization, evaluated in the vacuum of mass $m_1$ is
\bea\label{eq:T00}
 \langle T_{00} \rangle_1 &=& \frac{1}{2}\langle \pi(x)\pi(y) \rangle_1 + \frac{1}{2} \langle \nabla_{x}\phi(x)\cdot \nabla_y \phi(y) \rangle_1 + \frac{1}{2} m_0^2  \langle  \phi(x)\phi(y) \rangle_1\nonumber\\
 &&\hspace{1cm}-\xi_c \langle\nabla^2_x \phi(x)\phi(y)+\phi(x)\nabla^2_y \phi(y)+2 \nabla_x \phi(x)\nabla_y \phi(y) \rangle_1\,.
\eea
Here $\langle \ldots \rangle_1$ means that the expectation value is taken in the state specified by ${\cal T}_1$.
It is important to take first $x_0=y_0$ and then the limit $|\vec{x}-\vec{y}|\rightarrow 0$. Note that the last term, giving the improvement term contribution to $T_{00}$, vanishes identically by translation invariance.  

Before proceeding to the calculation, let us see how (\ref{mudu}) works out in this case. If we set $m_0=0$, $T_{\mu\nu}$ is an explicitly traceless operator, that we should write in terms of $\phi(x)$ and $\pi(x)$ for $x$ in the spatial surface $x_0=0$ before proceeding to evaluate expectation values in the theory of mass $m_1$. This needs the massless equations of motion for $d>2$ and the $i,j$ components of the stress tensor, since these contain $\partial_0^2 \phi$ in the improvement term. Once this is done, the operator $T_{\mu\nu}$ is explicitly traceless.   
  Using the isotropy of the spatial surface, we have
\be
\langle T_{ij} \rangle_1 = \frac{\langle T_{00}\rangle_1}{d-1}\,\delta_{ij}\,.
\ee
Comparing with (\ref{mudu}), we then have $\eta_\mu = \delta_{\mu0}$, and $k = \frac{d}{d-1} \langle T_{00}\rangle$. This illustrates how the dependence on the Cauchy surface appears for the simple case of a free scalar.

Given (\ref{eq:T00}), we can now evaluate $\Delta \langle T_{00}\rangle = \langle T_{00}\rangle_1-\langle T_{00}\rangle_0 $.
In $d=2$ and with $x_0=0$ we have
\begin{eqnarray}
\Delta \langle T_{00}\rangle &=& \frac{1}{4 \pi} \left[  (m_0^2-m_1^2)K_0(m |\vec{x}|) - \frac{2 m_1}{|\vec{x}|} K_1(m |\vec{x}|) + 2 \frac{m_0}{|\vec{x}|}K_1(m_0 |\vec{x}|) \right] \nonumber\\
&=& \frac{1}{8 \pi} \left[ m_1^2 - m_0^2 + 2 m_0^2 \log(m_0/m_1)   \right]\label{miti}
\end{eqnarray}
in the limit $|\vec x| \to 0$. This function is positive for all $m_0$ and $m_1$, reaching a minimum of zero in $m_0$ as a function of $m_1$. As we have seen, for $m_0=0$ this positivity is necessary to have a positive relative entropy in the interval. For $m_0\neq 0$ the positivity of this quantity is still needed for positivity of relative entropy in Rindler space, where the modular Hamiltonian is still given in terms of $T_{00}$.

If instead of doing the point splitting on the $x_0=0$ surface we choose another spatial direction $x_0= \alpha x_1$, with $|\alpha|<1$, we can split the points along this line to find
\be\label{eq:pp1}
\Delta \langle T_{00}\rangle  = \frac{1}{8 \pi} \left[\frac{\alpha^2 +1}{\alpha^2 - 1} (m_0^2 - m_1^2) + 2 m_0^2 \log(m_0/m_1)   \right]
\ee
as the regulator vanishes. This is not positive for all the range of $m_0$, $m_1$. The reason is that in using a point splitting in a slanted direction we have made use of correlators of the ${\cal T}_1$ theory outside the Cauchy surface. Recalling these expectation values in different Cauchy surfaces belong to different states for the ${\cal T}_0$ theory, we are not able to justify positivity from relative entropy, and in fact positivity fails.     

Let us consider next a hard momentum cutoff. Since the Cauchy surface at $x_0=0$ distinguishes space and time, we will allow for two different cutoffs on momenta, $|p_0|<\Lambda_0$, $|\vec p\,| <\Lambda$. The physical limit corresponds to $\Lambda_0 \gg \Lambda$, so that we have a spatial lattice that propagates in a continuous time variable. For Lorentz (or euclidean) invariant quantities, the order in which the cutoffs are sent to infinity does not matter. One then usually chooses $\Lambda_0= \Lambda$ to be able to use euclidean invariance. Here, however, we will see that $\Lambda_0 \gg \Lambda$ and $\Lambda_0 \ll \Lambda$ give different results for $\Delta \langle T_{00}\rangle$. To simplify the formulas, consider $m_0=0, m_1=m$, and let us work in $d=2$. Fourier-transforming $T_{00}= \frac{1}{2} (\partial_0 \phi)^2+\frac{1}{2} (\partial_1 \phi)^2$ and rotating to euclidean signature, we have,
\be
\Delta \langle T_{00} \rangle =\frac{1}{2} \int_{-\Lambda_0}^{\Lambda_0} \frac{dp_0}{2\pi} \int_{-\Lambda}^{\Lambda} \frac{dp_1}{2\pi}\, (-p_0^2+p_1^2) \left(\frac{1}{p_0^2+p_1^2+m^2}-\frac{1}{p_0^2+p_1^2} \right)\,.
\ee
If $\Lambda_0 \gg \Lambda$, we perform first the integral over $p_0$, and can take $\Lambda_0 \to \infty$. The resulting integral over $p_1$ is then finite and agrees with the point-splitting result (\ref{eq:pp1}). If we instead take $\Lambda \gg \Lambda_0$ and integrate over $p_1$ first, the final result has opposite sign,
\be
\Delta \langle T_{00}\rangle= - \frac{m^2}{8\pi}\,,
\ee
which is not physical for the energy.

As a last example for $d=2$ we can compute $T_{00}$ in a lattice. We use first neighbors
\be
T_{00}(i)=\frac{1}{2}(\pi_i^2+(\phi_i-\phi_{i-1})^2+m_0^2 \phi_i^2)\,,
\ee
and the lattice correlators in the fundamental state 
\bea
\langle \phi_k \phi_l\rangle= \frac{1}{4\pi}\int_{-\pi}^\pi dx\, \frac{e^{i x (l-k)}}{\sqrt{m_1^2+2 \cos(x)}}   \,, \\
\langle \pi_k \pi_k\rangle = \frac{1}{4\pi}\int_{-\pi}^\pi dx\, e^{i x (l-k)}\,\sqrt{m_1^2+2 \cos(x)}   \,.
\eea
The result for $\Delta \langle T_{00} \rangle$ in the limit of small lattice spacing coincides with (\ref{miti}).

The results for higher dimensions can be similarly calculated. Using point splitting we obtain
\begin{eqnarray}
d=2 &&\Delta \langle T_{00}\rangle = \frac{1}{8 \pi} \left[ m_1^2 - m_0^2 + 2 m_0^2 \log(m_0/m_1)   \right]\,, \nonumber\\
d=3 &&\Delta \langle T_{00}\rangle = \frac{1}{24 \pi} (m_1-m_0)^2 (m_1 + 2 m_0)\,, \nonumber\\
d=4 &&\Delta \langle T_{00}\rangle = -\frac{(m_0^2-m_1^2)^2}{32 \pi^2}\log(\epsilon)\,,\nonumber\\
d=5 &&\Delta \langle T_{00}\rangle = \frac{(m_1^2-m_0^2)^2}{64 \pi^2 \epsilon}\,, \\
d=6 &&\Delta \langle T_{00}\rangle = \frac{(m_1^2-m_0^2)^2}{64 \pi^3 \epsilon^2}\,,\nonumber\\
&&\ldots\nonumber\\
d=d && \Delta \langle T_{00}\rangle \propto \frac{(m_1^2-m_0^2)^2}{ \epsilon^{d-4}}\nonumber\,.
\end{eqnarray}
We see these are all positive for all $m_0$, $m_1$, as expected. The perturbation of the Hamiltonian due to a mass has dimension $\Delta=d-2$, with coupling constant $m_1^2$. These results match the expectations of a finite $\Delta T_{\mu\nu}$ for $\Delta<d/2$, which gives $d<4$. In fact for the finite cases $d=2,3$ we obtain the same results with other regularizations, such as a lattice. For the divergent cases $d\ge 4$ the results also match the expectations from conformal perturbation theory (for $m_0=0$), that is, $\Delta \langle T_{00}\rangle\sim g^2/\epsilon^{2 \Delta-d}=m_1^4/\epsilon^{d-4}$. We obtain similar results for free fermions. However, for fermions $\Delta \langle T_{00}\rangle\sim m^2/\epsilon^{d-2}$ diverges in all dimensions, as corresponds to $\Delta=d-1$. Nevertheless, on the null surface the relative entropy is finite for $d=2,3,4,5$ for scalars (up to a subtlety that we will address next), and $d=2,3$ for fermions.

%%%%%%%%%%%%%%%%%%%%%%%%%%%%%%%%%%%%%%%%%%%%%%
%%%%%%%%%%%%%%%%%%%%%%%%%%%%%%%%%%%%%%%%%%%%%%
\subsection{Boundary term in the modular Hamiltonian}\label{subsec:Hbdry}

In the power-counting classification of Sec. \ref{subsec:regimes} there is a subtle point for free scalars. These have divergent $\Delta S$ for $d\ge 4$, and the dimension of the relevant perturbation $m^2\phi^2$ is $\Delta=(d-2)$. Hence they violate the standard counting which would produce divergences for $\Delta=d-2\ge (d+2)/2$ and then $d\ge 6$. We will now see that in fact this divergence cancels out (at least for $d=4$) from the relative entropy because of an additional boundary term in the modular Hamiltonian \cite{Lee:2014zaa, Lewkowycz:2014jia, Herzog:2014fra, Casini:2014yca}.

The free scalar theory contains a subtlety that is generically absent from more general flows: the improvement term in the conformal stress tensor (\ref{eq:Tmunu-scalar}). The modular Hamiltonian in Rindler space is constructed with the canonical stress tensor rather than the conformal one. The sphere modular Hamiltonian comes from the Rindler one by a conformal transformation, and we have to use the conformal tensor. Adding the improvement term to the canonical tensor gives an additional boundary term proportional to $\phi^2$~\cite{Casini:2014yca},
\be
\langle H_{bdry}\rangle=2\pi \xi_c \int d\sigma\,\langle \phi^2 \rangle\,,\label{dhj}
\ee
where the integral is over the boundary of the spherical entangling surface. This term does not change with Cauchy surface and subsists in the null limit. Hence we have to add (\ref{dhj}) to $-\Delta S$ to obtain the relative entropy.

The expectation value of $\phi^2$ on the state corresponding to the massive theory is
\be
\langle \phi^2(x) \rangle= \int \frac{d^d k}{(2\pi)^d}\,\frac{1}{k^2+m^2}\,.
\ee
This is finite for $d=2,3$ and divergent for $d\ge 4$. For $d\ge 4$ we can still get a universal part that is the finite term for $d$ odd and the logarithmic term for $d$ even. These universal pieces agree when computed using different regularizations, for example dimensional regularization, heat kernel, or point splitting.  
The general result for the universal term using dimensional regularization writes
\be
\langle \phi^2\rangle= \frac{2\pi^{d/2}}{(2\pi)^d \Gamma(d/2)}\,\frac{\pi}{2 \sin \frac{\pi d}{2}} (m^2)^{d/2-1}\,,
\ee
where an expansion in $d$ is assumed for even dimensions to get the logarithmic term. 
The boundary contribution corresponding to this universal part then reads
\be
\langle H_{bdry}\rangle =2\pi \frac{d-2}{4(d-1)} \frac{2\pi^{d/2}}{(2\pi)^d \Gamma(d/2)}\,\frac{\pi}{2 \sin \frac{\pi d}{2}} (m^2)^{d/2-1} A\,,
\ee
where $A$ is the area of the $d-2$ dimensional  spherical entangling surface.

Let us compute the relative entropy in the infrared for different dimensions. The universal pieces of the entropy are given by
\be
\Delta S= (-1)^{(d-1)/2} \frac{\pi}{3\,2^d \, \pi^{d/2-1}\, \Gamma[d/2]} m^{d-2}\,A\,,
\ee 
for $d$ odd, and 
\be
\Delta S= (-1)^{d/2} \frac{1}{3\,2^{d-1}\, \pi^{d/2-1} \Gamma[d/2]} \log(m \epsilon) m^{d-2}\,A\,,
\ee
for $d$ even \cite{Hertzberg:2010uv}.

Therefore, for $d=3$ we have $\Delta \langle {\cal H} \rangle_{\Sigma_{\text{null}}} =-\frac{1}{16} m A$ and $\Delta S= -\frac{1}{12} m A$. Note $\Delta \langle {\cal H} \rangle_{\Sigma_{\text{null}}}$, coming exclusively from the boundary term, is negative. However, the relative entropy is still positive, with a smaller area term than $-\Delta S$,
\be
S_{\textrm{rel}}=\frac{1}{48} m A\,.
\ee

For $d=4$ both $\Delta S$ and $\Delta\langle H_{bdry}\rangle$ are logarithmically divergent. However, these logarithmic terms agree,
\be
\Delta \langle H_{bdry}\rangle =\Delta S = \frac{1}{24\pi}\,\log (m\epsilon)\,  m^{2} A \,.
\ee
Therefore, these divergences cancel out of the relative entropy. Thinking in the relative entropy on the null surface as a form of regularization of the entropy, this restores the validity of the counting argument in Sec. \ref{subsec:regimes} for free scalars. Once the divergent parts cancel, there must remain a finite term  proportional to $m^2 A$ for $S_{\textrm{rel}}$ in $d=4$. To get this area term requires using the same cutoff for the entropy and $\Delta\langle\phi^2 \rangle$. It would be interesting in the future to calculate $S_\text{rel}$ explicitly in terms of a physical cutoff. Here we will simply assume that the power-counting analysis of Sec. \ref{subsec:regimes} becomes valid due to cancellations between $\Delta \langle H_{bdry}\rangle$ and $\Delta S$.

For $d=5$ both $\Delta S\sim \epsilon^{-1}$ and $\Delta\langle\phi^2 \rangle\sim \epsilon^{-1}$. If the boundary term generally restores the counting of divergences for the scalar, we should also have finite relative entropy in $d=5$. This would mean that the leading divergences cancel, and we end up with the universal pieces. For these we have $\Delta S\sim m^3/(64 \pi)$ and $\Delta\langle {\cal H}\rangle_{\Sigma_{\text{null}}} \sim m^3/(72\pi)$. Again the result is positive,
\be
S_{\textrm{rel}}=\frac{1}{576 \pi} m^3 A\,.
\ee

Finally, for $d=6$ the naive counting gives a logarithmically divergent $S_{\textrm{rel}}$. If all higher powers cancel, we have from the universal parts $\Delta S\sim -1/(192 \pi^2) \log(m\epsilon) m^4 A$ and $\Delta\langle {\cal H}\rangle_{\Sigma_{\text{null}}} \sim -1/(160 \pi^2) \log(m\epsilon) m^4 A$. This gives the divergent, though positive result
\be
S_{\textrm{rel}}=-\frac{1}{960 \pi^2} \log(m\epsilon) m^4 A\,.
\ee 
For $d\ge 7$ the combination of the universal parts is not positive, which is consistent with the relative entropy having leading divergent non universal terms that compensate for the sign.

\bibliography{EE}{}
\bibliographystyle{utphys}

%%%%%%%%%%%%%%%%%%%%%%%%%%%%%%%%%
\end{document}